\documentclass[article,nofootinbib]{revtex4}

\usepackage{graphicx}
\usepackage{epsfig}
\usepackage{fancyhdr}
\usepackage{empheq}
\usepackage{mathbbol}
\usepackage{wasysym}
\usepackage{bbm}
\usepackage{bm}
\usepackage{color}
\usepackage{amssymb,amsmath}
\usepackage{epstopdf}
\usepackage{float}
\usepackage{color}
\usepackage{subfig}

\graphicspath{{./png/}}

\newcommand{\be}{\begin{equation}}
\newcommand{\ee}{\end{equation}}
\newcommand{\bea}{\begin{eqnarray}}
\newcommand{\eea}{\end{eqnarray}}

\begin{document}

\title{New Spectroscopy with Charm
and Beauty Multiquark States}
\author{Luciano MAIANI}
\address{ Dipartimento di Fisica, Universit\`a di Roma "La Sapienza" , Piazzale A Moro 5, Roma, I-00185, Italy; \\
CERN, PH-TH, 1211 Gen\`eve 23, Switzerland.\\ \\
Talk given at the Erice School of Subnuclear Physic, June 30, 2013, Ettore Majorana Foundation and Center for Science and Culture, Erice, Italy.}

\begin{abstract}
{\bf Abstract.}
Exotic charmonium and bottonomium resonances recently discovered are discussed and interpreted as diquark-antidiquark states containing a pair of charm quarks and a pair of light, up and down, quarks. Successes, shortcomings and predictions of the model are illustrated.

\noindent
--------------------

\noindent
\end{abstract}

\maketitle

\thispagestyle{fancy}

\begin{figure}[t]
\begin{center}
\includegraphics[scale=.85]{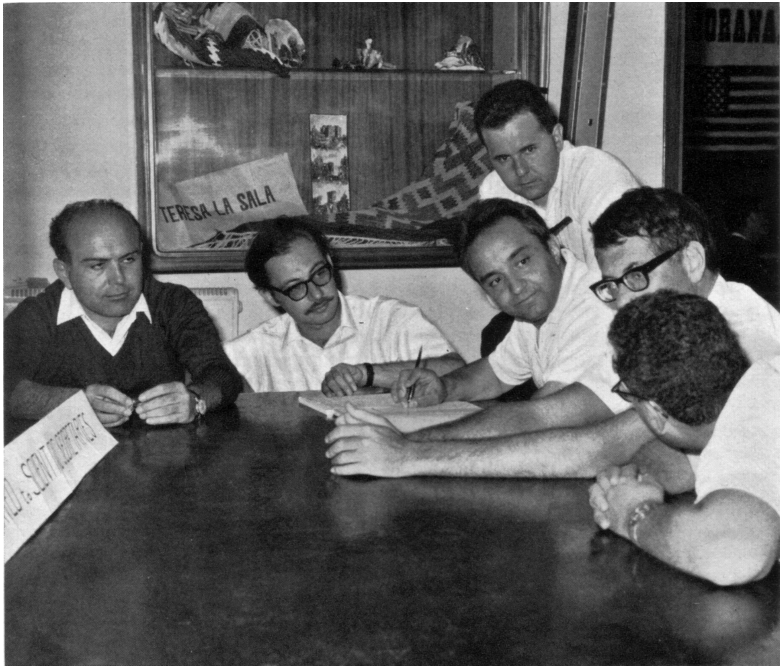}       
\caption{\label{fig:nicetal}  {\footnotesize   Making the Standard Model, Erice School in Subnulear Physics, 1967. Going around the table from left: Bruno Zumino, Sidney Coleman, Nino Zichichi, Nicola Cabibbo, Sheldon Glashow and Murray Gell-Mann. CERN Courier, Ref.~\cite{courier}}}
\end{center}
\end{figure}

\section{Introduction} 
For a long time, we lived happily with the Gell-Mann Zweig paradigm:
\bea
&&{\rm mesons} =q\bar q\notag \\
&&{\rm  baryons} = qqq
\eea

Paradigm rested on the absence of $I=2,~\pi\pi$ resonances and of $S>0$ baryons.

The case had to be revisited, because the lowest lying, scalar mesons - the $\sigma$(600), $f_0$(980), and $a_0$(980)-do not fit in the picture.
It was also found, much later, that {\it new ÒcharmoniumÓ} states seen by Belle and BaBar, do not fit the charmonium picture at all.

The discovery of $\Theta^+,~ S=+1$ baryon (a pentaquark ?), seemed to make the case...but the first observations have not been confirmed.

The situation has considerably progressed since 2007, when I gave two lectures on tetraquarks in Erice~\cite{maia07}.  The evidence for charmonium and bottomonium states made by four valence quarks is now established, with the discovery of {\it charged charmonia and bottomonia} and the wounderful experiments performed by BELLE, BaBar,  CDF and D0, BES, CLEO and now CMS and LHCb. Some of the new particles have been observed at Tevatron and LHC, data on decay modes and masses of the new particles are accumulating and theoretical models explored. 

We are still in a growing stage, with contrasting evidences coming and going. The situation looks quite similar to the one prevaling for meson spectroscopy in the sixties, when the Erice School started and the Standard Model was on the making, see the wonderful photo in Fig.~\ref{fig:nicetal}. 
From the report of that school~\cite{courier}, we learn that: 

{\it Professor Hughes from the University of Glasgow, UK, presented a review of the current situation regarding the meson resonances....Then the data ond masses and quantum numbers of the bosons were examined in detail in order to determine {\bf what is certain, what remains to be confirmed and what should probably be forgotten}} (emphasis added).

There will be, no doubt,  many data and theories about the new resonances that will have to be forgotten. But non conventional hadrons are going to stay with us, giving important clues  to the understanding of strong interactions dynamics.

\begin{figure}[t]
\begin{center}
\includegraphics[scale=.350]{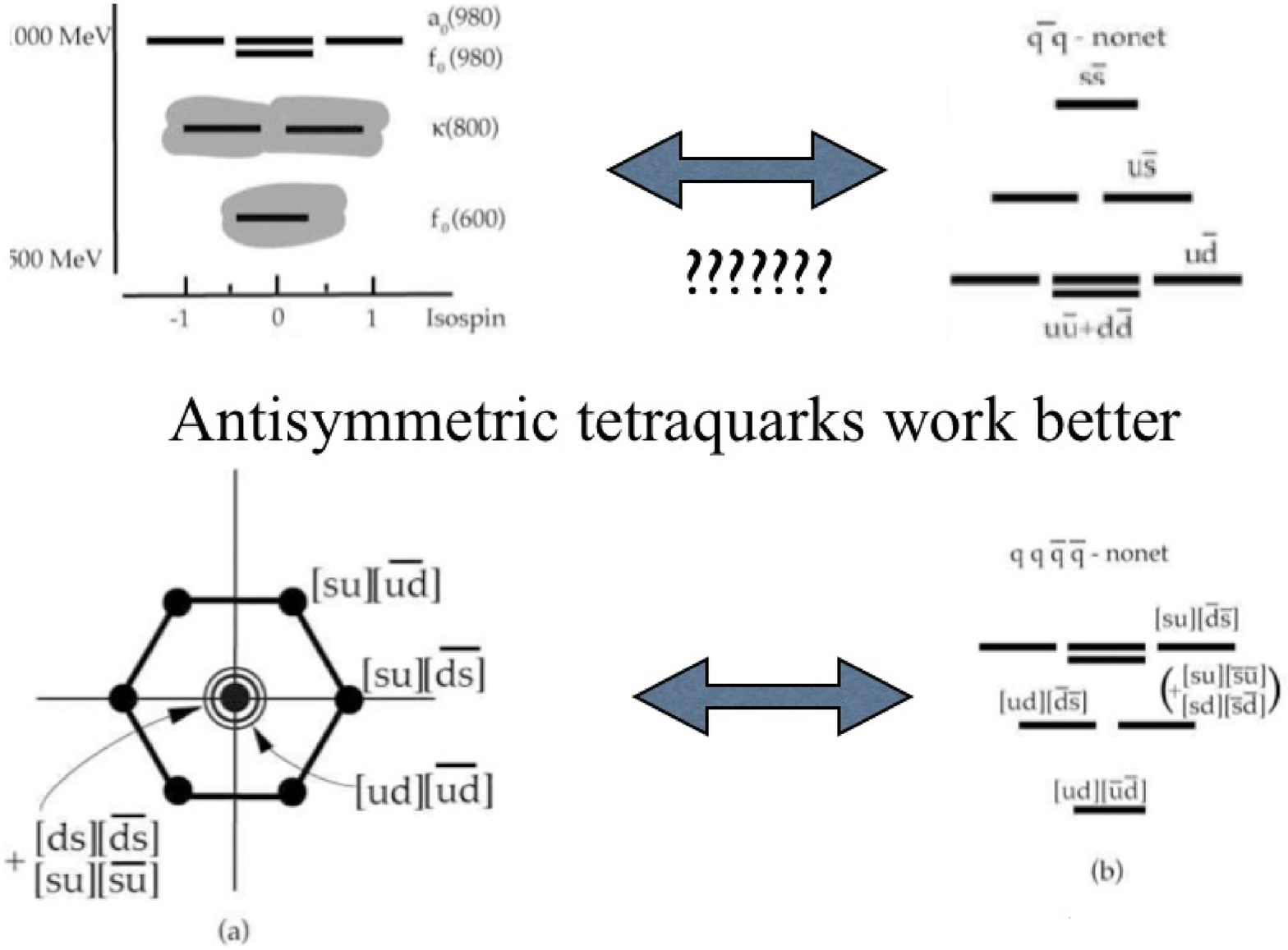}       
\caption{\label{spectrum}  {\footnotesize  .}}
\end{center}
\end{figure}

\section{Light scalar mesons}

A light scalar meson nonet, with masses below 1 GeV, has been long suspected or claimed, but only recently established. The spectrum of these particles is completely inverted with respect to the the spectrum of a standard meson nonet, e.g. the vector meson nonet. This is exemplified in Fig.~\ref{spectrum}. The isospin one states are degenerate to an isosinglet, which is appropriate if they are made by the light quarks. However, the $a_0(980)$, $I =1$, and $f_0(980)$, $I=0$, are heavier than both the strange meson $\kappa(800)$ and the other singlet $f_0(600)$. One could explain the low mass of $f_0(600)$ as being pushed down by a mixing with a higher lying glueball, as assumed e.g. in~\cite{fritzsch}, but this would not explain why the strange meson is lower than the isovector.

As noted by Jaffe long ago~\cite{jaffe}, antisymmetric diquarks work better. 
A pair of light quarks in fully antisymmetric state could have sufficent attraction to make the basic unit of a new series of hadrons (Jaffe~\cite{jaffe}, Jaffe and Wilcezck~\cite{jaffewil}). For light quarks, we denote by $[qq]$,~($q=u,d$), the state:
\bea
&& [qq]: 
~~ {\rm color} = \bar 3,~{\rm SU(3)~flavour} = \bar 3,~{\rm orbital~angular~momentum}~L=0,~{\rm total~spin}=0
\label{asdiquark}
\eea
A color singlet, diquark-antidiquark pair would make a flavor nonet with a composition such as to reproduce perfectly the ordering of the light scalars, with the $I=1,0$ composed as $[sq][\bar s\bar q^\prime]$ being the heaviest components, as shown in the lowest part of Fig.~\ref{spectrum}. 

\begin{figure}[t]
\begin{center}
\includegraphics[scale=.350]{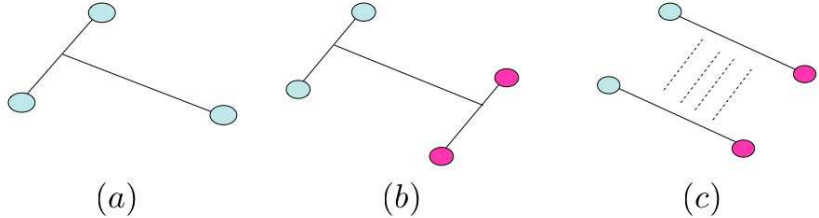}       
\caption{\label{fig:strings}  {\footnotesize  Diquark binding by color forces: (a) in baryons, (b) in tetraquarks; by contrast, meson-molecules (c) are bound by residual color forces due to normal meson exchange.}}
\end{center}
\end{figure}

\section{{\bf Color forces, decays and heavy diquarks.}}

Diquarks needs to combine with other colored objects to form color neutral hadrons. Binding to another quark, with Y-shaped string, a baryon is formed, Fig.~\ref{fig:strings}(a). Diquarks bound to an anti-diquark, with an H-shaped string, form a new kind of mesons, Fig.~\ref{fig:strings}(b). We call these states {\it tetraquarks} for brief. Tetraquarks in QCD at large $N$ have been considered recently by S. Weinberg~\cite{Weinberg:2013cfa}.

We shall consider also diquarks containing one heavy quark and the corresponding {\it hidden charm or beauty tetraquarks} with composition: $[cq][\bar c \bar q^\prime]$. The spin-spin interactions of the heavy quarks are suppressed in QCD, as exemplified by the small splitting of  charmed and beauty pseudoscalar and vector mesons. Thus it is reasonable to imagine that diquarks with one heavy quark may be formed irrespectively of the value of the spin of the pair:
\be
S_{cq}=0, 1
\ee 

Hidden charm/beauty tetraquarks should make full flavor multiplets corresponding to  the composition of the diquark and antidiquark total spin, with radial and orbital excitations.   

\begin{figure}[t]
\begin{center}
\includegraphics[scale=.50]{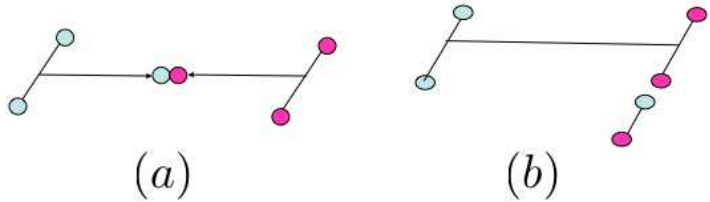}       
\caption{\label{fig:break}  {\footnotesize  (a) Tetraquark decay in Baryon-Antibaryon pair, (b) de-excitation in a lower tetraquark.}}
\end{center}
\end{figure}

A different configuration of four quarks is that of a meson-meson molecules, Fig.~\ref{fig:strings}(c). Such hypothetical states would be bound by the residual  forces of QCD, similar to the nuclear forces which act upon color singlets due to light mesons exchange. 

The question for molecules is wether these forces are at all able to bind them. The residual forces will not be flavor blind and, most likely, will give rise to few states, in each channel, possibly with incomplete SU(3) and even SU(2) multiplets

The string topology of tetraquarks is related to the baryon-antibaryon channel~\cite{veneziarossi}. As shown in Fig.~\ref{fig:break} (a), by breaking the string one gets a $B\bar B$ pair:
\be
[qq][\bar q \bar q]\to B \bar B
\ee
In many cases, such decays are forbidden by energy conservation. A tetraquark may decay into a lighter one with the same composition by emission one meson,~Fig.~\ref{fig:break} (b):
\bea
&&([qq][\bar q \bar q])_A\to ([qq][\bar q \bar q] )_B+ {\rm meson (s)}
\label{deexcit}
\eea
or dissociate directly into a $J/\Psi$ or a $\Upsilon$ and one or more light mesons, if diquarks contain one charm or one beauty quark.

\paragraph*{\bf {Instanton effects.}}

Instantons produce an effective lagrangian corresponding to a six-quark vertex~\cite{gerardt013}. For tetraquarks, they are able to produce important effects , which were identified precisely in Erice in 2007~\cite{gerardt07}, see also~\cite{schechter}:

(a) mixing of  the lightest tetraquark scalars with the lowest $q\bar q$ mesons, identified with the multiplet made by $a_0$(1474) (I=1), $K_0$(1412), (I=1/2), and three isosinglets: $f_0$(1370), $f_0$(1507) and $f_0$(1714) (one could be a glueball);

(b) a mechanism for the decay $f_0$(980)$\to \pi^+\pi^-$.

\begin{figure}[t]
\begin{center}
\includegraphics[scale=.50]{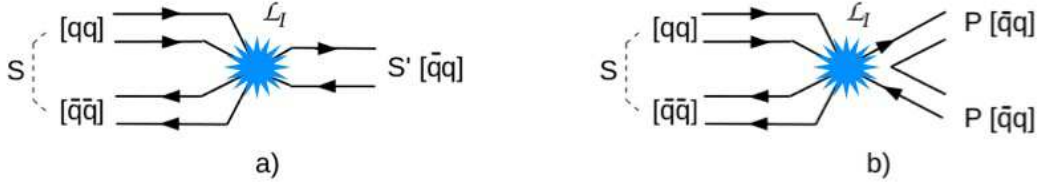}       
\caption{\label{fig:instant}  {\footnotesize  (a) Mixing of the light tetraquark with a $q\bar q$ meson, (b) mechanism for the $f_0\to 2\pi$ decay.}}
\end{center}
\end{figure}

\section{Conventional and less conventional Quarkonia}
The accuracy with which the spectra of $Q \bar Q$ states (Q=c, b) can be predicted and measured makes it possible to discover new states Òby differenceÓ~\cite{pdg, QWG}.
Terminology of $Q \bar Q$ states in S and P wave is as follows.

A state with:
\be
{\rm spin}=S,~{\rm orbital~ang.~mom.}=L,~{\rm total~ang.~mom.}=J,~{\rm radial~excitation}=n\notag 
\ee
is denoted by: $n~^{2S+1}L_J$.

Lowest lying states of charmonium are given special notations:
\bea
&&\eta_c(1S)=1~^1S_0;~J/\Psi=1~^3S_1\\
&&h_c(1P) =1~^1P_1,~\chi_c(1P) =1~^3P_J 
\eea

\begin{figure}[t]
\begin{center}
\includegraphics[scale=.50]{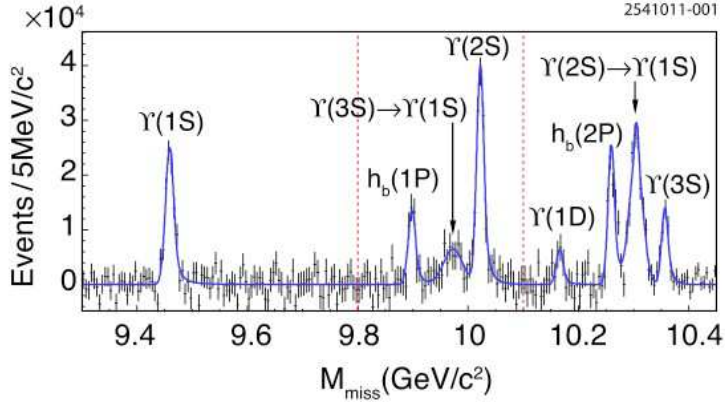}       
\caption{\label{fig:2pispectr}  {\footnotesize  Mass recoiling against $\pi^+\pi^-$ in $e^+ e^-$ collisions. Spectrum from BELLE reported in Ref.~\cite{pdg}.}}
\end{center}
\end{figure}

The spectrum reported in Fig.~\ref{fig:2pispectr} gives an idea of the complexity of bottomonium lines, with resonances attributed to levels by comparing the measured masses and decay branching ratios to the theoretical calculations. 
The situation is reminiscent of the XIX century determination of the elemental composition from the spectral lines of stars. 
And, similar to this example, sometimes {\it unanticipated lines} appear. 

\begin{figure}[t]
\begin{center}
\includegraphics[scale=.50]{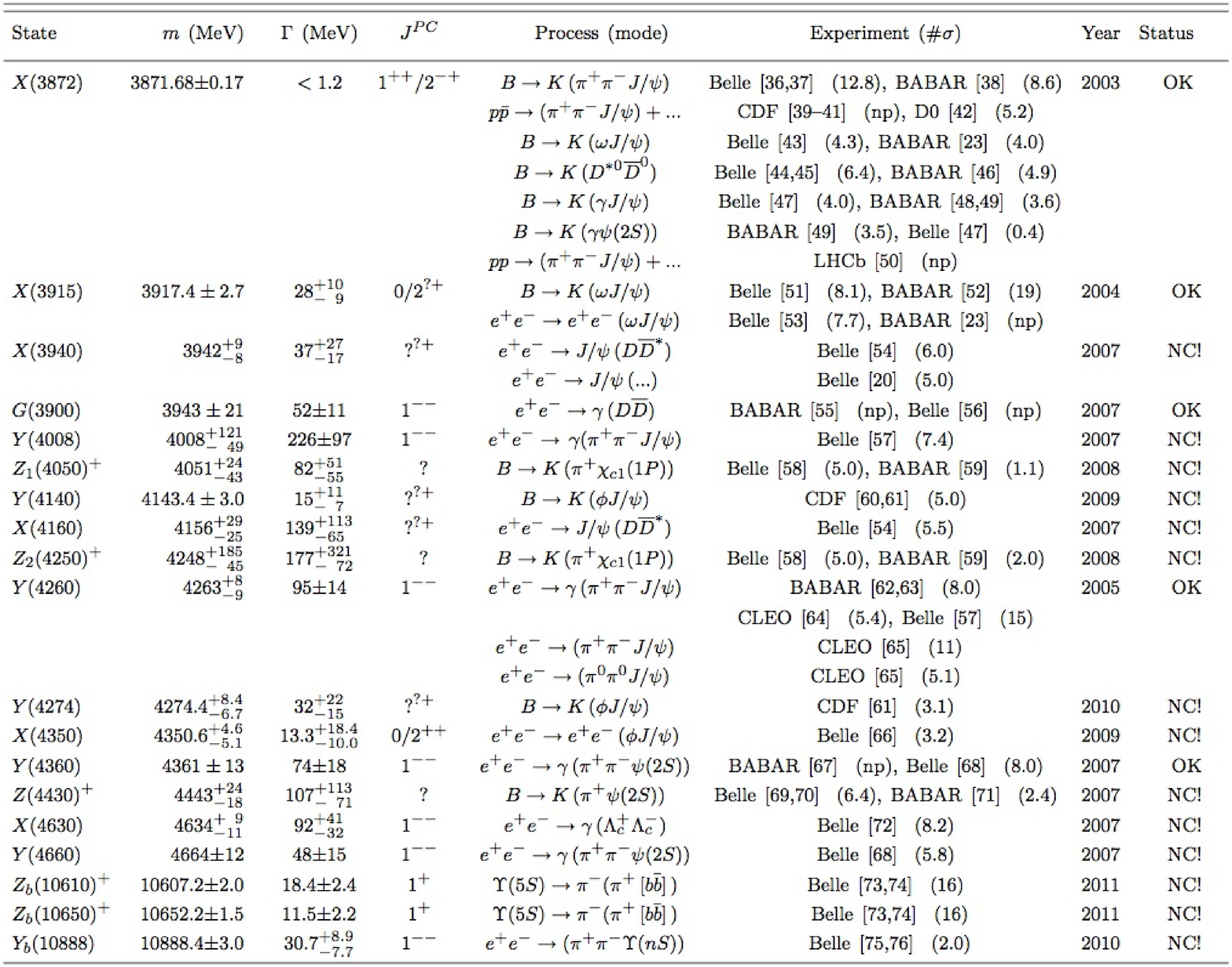}       
\caption{\label{fig:unantic}  {\footnotesize  Unanticipated hidden charm and beauty states from Ref.~\cite{pdg}.}}
\end{center}
\end{figure}

A recent list of unaccounted for states is given in Refs.~\cite{pdg,QWG} and reproduced here\footnote{In a recent paper by  T. Friedmann~\cite{Friedmann:2009mx} the existence of radial excitations is challenged and  the states usually classified as such are interpreted as tetraquark states. The list of "unanticipated states" is therefore considerably increased, to include even the well known $\psi^\prime$, usually classified as $\psi(2~^3S_1)$ and now included in the class of $Y$ states. I shall  not follow this line, but note that this interpretation would put the observed decay $\psi^\prime \to J/\Psi~\pi^+ \pi^-$ in line with the decays of $X(3872)$ or $Z_c(3900)$ into $J/\Psi +$~meson(s).}, Fig.~\ref{fig:unantic}.

The established convention for these states is that $X$ are neutral states which have positive parity if their decay (e.g. into $J/\Psi+\rho$) happens to be in S-wave, $Y$ are neutral $J^{PC}=1^{--}$ resonances seen in $e^+e^-$ annihilation, and $Z^\pm$ are electrically charged states, again with positive parity for S-wave decay into $J/\Psi+\pi^\pm$.

Some of the new lines, e.g. $X(3872)$, are very close to some meson-meson threshold:
\bea
&& M[X(3872)] =3871.68\pm 0.17~{\rm MeV} \notag \\
&& M[X(3872)] - M(D^0+D^{*0})= - 0.12\pm 0.30~{\rm MeV}
\label{massdiff}
\eea  
but some are not, e.g. $Y(4260)$. 


\section{What is X(3872) made from?}

Several hypotheses are still on the market, waiting to be supported or excluded by conclusive experimental data. Here the most frequently considered interpretations.
\begin{itemize}
\item The {\bf tetraquark hypothesis} would give rise to an hadron whose components are held together by color forces, to make an hadron of normal size ($\approx 1$~fm). Variants of the four quark picture can be envisaged with diquarks bound in color $6$ or $q\bar q$ pairs in color $8$.

\item The {\bf hybrid option}, with a $c\bar c$ pairs in color $8$ bound to a {\it constituent gluon}, see e.g.~\cite{close}. This option, however, would not apply to charged states like $Z_c(4430)$ or $Z_b$ and $Z^\prime _b$. 

\item{\bf Molecule}: the small value of the difference in (\ref{massdiff}) is in favor of a loosely bound state of the two mesons, held together by forces similar to the nuclear forces. Perhaps too much loosely bound, since the radius estimated form the binding energy is definitely anomalously large. From the non-relativistic Schroedinger equation one finds:
\be
R \approx \frac{\hbar c}{\sqrt{2M_{red}E_{bind}}}=\frac{200~ {\rm MeV}\cdot{\rm  fm}}{20{\rm -}30~ {\rm MeV}}=7~{\rm to}~10~{\rm fm}
\ee
where $M_{red}$ and $E_{bind}$ denote the reduced mass and binding energy and the numerical value follows from~(\ref{massdiff}). 

\item {\bf Hadro-charmonium}. A charmonium state surrounded by a light quark cloud has been recently envisaged~\cite{hadroc}. One could have also excited charmonium in the core, which could explain why $Z(4430)$ decays into $\psi(2S)$. The wave function of hadro-charmonium should depend from the relative separation, $R$, of $c$ and $\bar c$. For large $R$, light quarks/antiquarks may be attracted between the heavy particles to neutralize color forces, thereby mimicking a molecular state. Thus hadro-charmonium would have a molecular component, which however should be dominant if we wish to mantain the relation with meson masses, reproducing the puzzle of the anomalously large radius. In addition, it is not clear what would stabilize the light quark/antiquark cloud, which in principle is present also in normal charmonia, and promote some of the cloud's components into valence quarks.
\end{itemize}

\begin{figure}[t]
\begin{center}
\includegraphics[scale=.45]{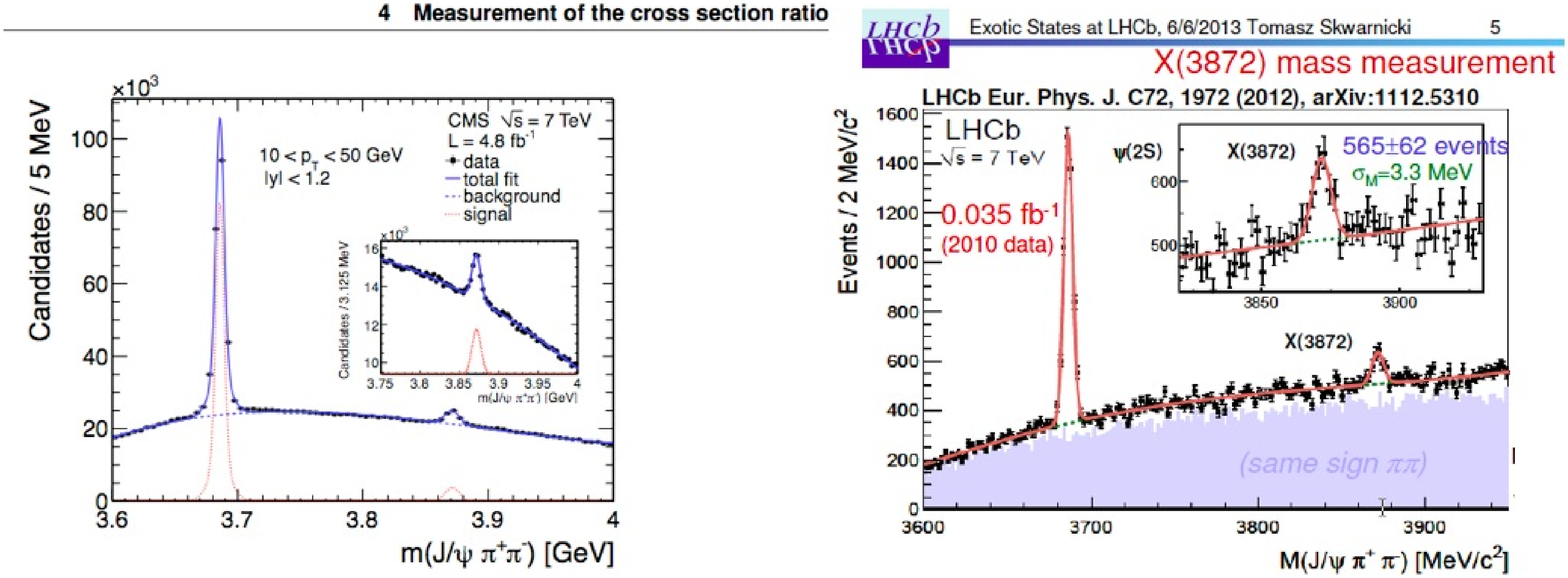}       
\caption{\label{fig:xatcoll}  {\footnotesize Observation of $X(3872$ at the LHC by CMS, left, and LHCb, right. The production of $\psi^\prime$ gives an idea of the cross section involved.}}
\end{center}
\end{figure}

\paragraph*{{\bf 
Production at Hadron Colliders.}}
Production of $X(3872)$ is seen at the Tevatron, by CDF, and LHC, by CMS and LHCb, at a lower but not so dissimilar level as charmonia, see Fig.~\ref{fig:xatcoll}. 

Using Pythia to estimate the probability to find a $D\bar D$ pair in the relevant phase space, factors of $10^{-2}$ with respect to the X(3872) cross section measured by CDF ($\approx  30$~nb) are found~\cite{bignamini}.

The observation at hadron colliders speaks against extended objects and in favor of compact hadrons, like tetraquarks. Several explanations have been advanced by the supporters of the molecular picture but, all in all, this remains to me a very weak point of this picture.

\paragraph*{{\bf Charged partners of the  $X(3872)$.}} Negative results for the search of a narrow, charged counterparts of the $X(3872)$ have been published by BABAR~\cite{babXchar} and BELLE~\cite{belXchar}. The pure $I=1$ hypothesis for $X(3872)$ is convincingly excluded, but there is still a non neglible parameter space for a mixed $I=1,0$ª situation. The possibility that the charged $X$ state may be wide enough to have escaped detection has been considered in Ref.~\cite{wide}

\section{Tetraquark picture of unexpected $X$ particles}

We consider the simplest  tetraquark configurations, in $S-wave$ and composition:
\be
[c q][\bar c \bar q^\prime]_{L=0},~(q,q^\prime=u,d)
\ee

The total spin of each diquark can assume the values $S=1, 0$ and the composition of the two spins gives rise to a supermultiplet of states with different total angular momentum, $J$, and positive parity, $P=+1$. 

For each value of $J$, light quark flavours imply that there are four states with isospin $I=1, 0$, analogous to the $f_0(980)$- $a_0(980)$ complex. 

We may also analyze the states in terms of the total spin of the quark-antiquark pairs, $S_{q\bar q^\prime}$ and $S_{c\bar c}$, which also may take the values $0,1$. An important quantum number is the Charge Conjugation parity of the neutral components, $C$. The value of $C$, in a state with given values of  $S_{q\bar q}$ and $S_{c\bar c}$ is:
\be
C=(-1)^{S_{q\bar q}+S_{c\bar c}}
\label{c-conj}
\ee
If isospin is exact, we may also consider the $G$ parity of each isospin multiplet:
\be
G=C(-1)^I
\ee
where $C$ refers to the neutral component and $I$ is the isospin of the multiplet. The two neutral mesons of a given $J^{PC}$ can mix (like in $\rho$-$\omega$ mixing) with $G$ and $I$ violated and $C$ conserved.


The predicted tetraquark multiplets are listed below (in bold the tentative attribution of observed states): 
\begin{itemize}
\item $J^{PC}=2^{++}$,~(2 multiplets)
\item $J{PC}=1^{++}$,~(1 multiplet): {\bf X}(3872); 
\item $J^{PC}=1^{+-}$,~ (2 multiplets): {\bf Z}$_c$ (3900), Z$^\prime_c$; {\bf Z}$_b$(10610), {\bf Z}$^\prime_b$(10650)
\item$J^{PC}=0^{++}$,~(2 multiplets)
\end{itemize}

In analogy with the non-relativistic costituent quark model, one may try to describe the mass spectrum of the various states, starting from a common diquark mass, $M_{[cq]}$ and mass differences due to spin-spin interactions. The corresponding Hamiltonian is~\cite{hiddenc}:
\be
H=2M_{[cq]}+2\sum_{i<j}\kappa_{ij}({\vec s}_i\cdot {\vec s}_j)
\label{spin-spin}
\ee

The spin-spin interaction coefficients  can be in part related to the baryon and meson spectrum (e.g. $\rho-\pi$, $\Delta-N$, etc., mass differences). 

The inter-diquark coupling, $(\kappa_{cq})_{\bar 3}$ is directly related to charmed baryon mass differences. Critical couplings are the extra-diquark couplings, $\kappa_{q\bar q}$, $\kappa_{c\bar c}$ and $\kappa_{c\bar q}$. From the meson spectrum we obtain the corresponding couplings for the quark pairs in color singlets, while in the case at hand these pairs are in a superposition of color singlet and octet. Moreover, the spin-spin interaction is proportional to the overlap probability of the pair. We may relate the latter couplings to the meson ones, assuming~\cite{hiddenc}:
\begin{itemize}
\item one-gluon exchange: this means that the interaction constant $\kappa_{ij}$ for a pair in a given color state $c$ is proportional to the value of the Casimir operator in the color state:
\be
(\kappa_{ij})_c \propto \;<\sum_A  \frac{\lambda_i^A}{2}\frac{\lambda_j^A}{2}>_c
\ee
\item same $|\psi(0)|^2$ as in the mesons
\end{itemize}
The result is:
\be
\kappa_{q\bar q}=\frac{1}{4} (\kappa_{q\bar q})_0
\ee
and the numerical values~\cite{hiddenc} are given in the Table.

  \begin{table}[hb]%
\label{sscoup}
\centering
\begin{tabular}{|lc|l|clc|l|cc|clc|c|}
\hline
$(\kappa_{cq})_{\bar 3}$ && $ \kappa_{q\bar q}$ && $\kappa_{c\bar c}$ && $\kappa_{c\bar q}$ && $(\kappa_{bq})_{\bar 3}$ && $\kappa_{b\bar q}$ && $\kappa_{b\bar b}$ \\
\hline
$22$ && $79$ && $15$ && $18$ && $6$ && $6$ && $9$\\
\hline 
\end{tabular}
\caption{{\footnotesize Spin-spin couplings (in MeV) for tetraquarks with hidden charm and beauty derived from the meson and baryon spectrum. See text for the additional assumptions.}
} 
\label{tab:couplings} 
\end{table}

The constituent quark masses derived from meson and baryon spectrum, are:
\be
M_q=305,~M_c=1670,~{\rm (in~ MeV)}
\ee
while the diquark mass, fixed by the $X(3872)$ mass, is:
\be
M_{[cq]} = 1933~{\rm MeV}
\label{cdiq}
\ee
that is, additivity is satisfied within few tens of MeV. 

\begin{figure}[t]
\begin{center}
\includegraphics[scale=.6]{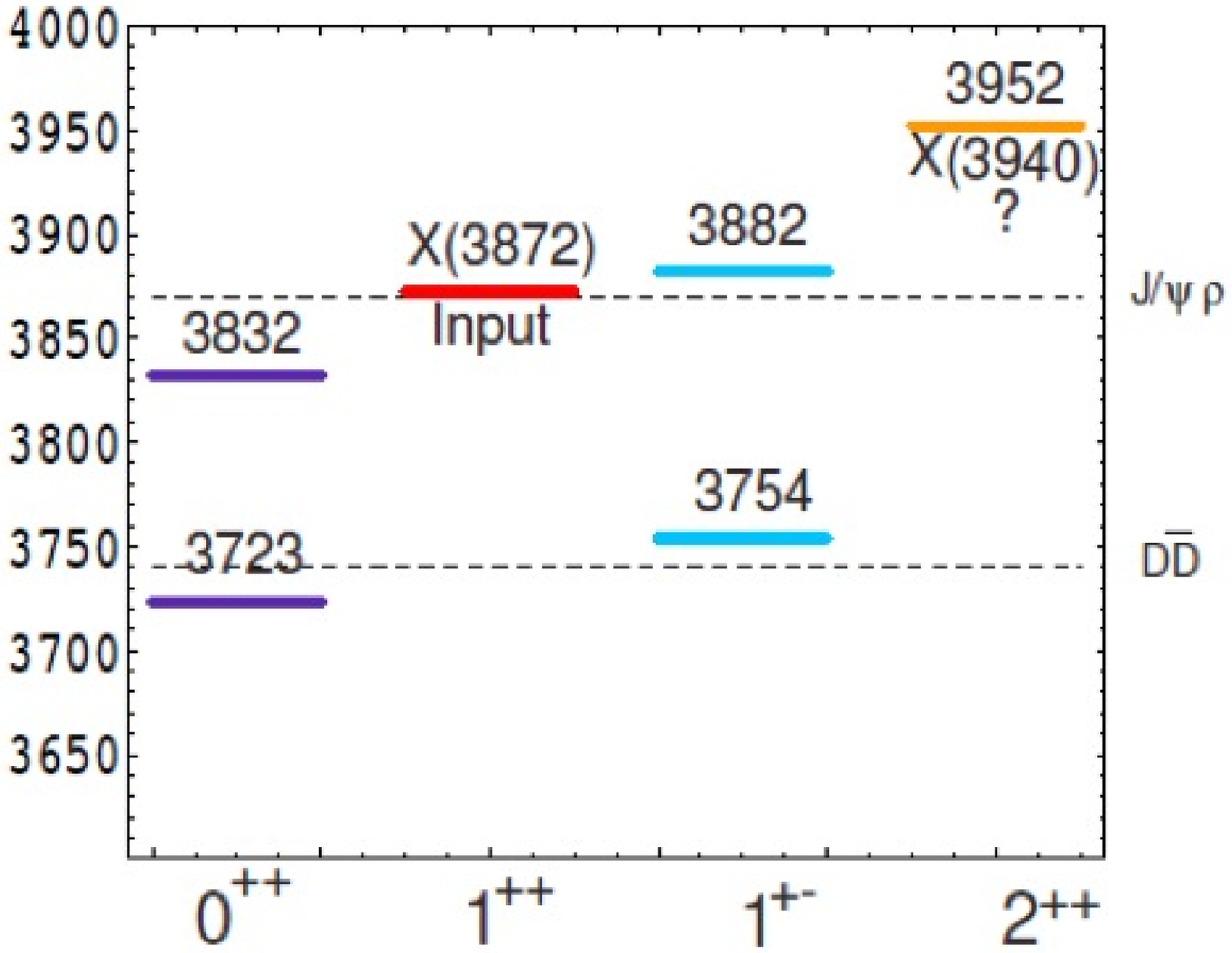}       
\caption{\label{spectrum2}  {\footnotesize .}}
\end{center}
\end{figure}

The mass spectrum of hidden charm X and Z particles from spin-spin interactions thus obtained is shown in Fig.~\ref{spectrum2}.

Tetraquarks with hidden beauty have been considered by Ali and collaborators, Ref.~\cite{zb4q}. The $[bq]$ diquark mass was derived from (\ref{cdiq}) by adding the beauty-charm constituent quark mass difference:
\bea
&&\Delta_{bc}=M_b-M_c=3335\notag \\
&& M_{[bq]}=M_{[cq]}+\Delta_{bc} = 5267
\eea

The mass difference of $Z_b$(10610) and Z$^\prime_b$(10650) agrees with the $B^*$-$B$ mass difference, as expected in the molecule model, see~\cite{zbmol}:
\bea
&&M(Z_b)-M(Z^\prime_b)=45~{\rm MeV}\notag \\
&& M(B^*)-M(B)=45.78 \pm 0.35~{\rm MeV}
\eea

Tetraquark model tends to predict larger splitting, depending however on the value of the poorly known $\kappa_{q\bar q}$, see~\cite{zb4q}.

\paragraph*{{\bf The $J^{PG}=1^{ +, \pm}$ charmonium tetraquarks}.}

We consider isotriplets only. There are three states with different charge conjugation properties of the neutral component and with the indicated distribution of the diquark total spin: 
\bea
&&C=+1:\; \frac{|1,0\rangle+|0,1\rangle}{\sqrt{2}};\; \; (3872~\text{MeV},\;{\rm input})\label{cplus}\\
&& C=-1:\;\frac{|1,0\rangle-|0,1\rangle}{\sqrt{2}};\; \; (3882~\text{MeV},\;{\rm computed} )\nonumber \\
&& C=-1: \;|1,1\rangle_{J=1};\; \; (3755~\text{MeV},\;{\rm computed} )
\label{cminus}
\eea
The state $1^{++}$ is a mass eigenstate, while states with $C=-1$ are mixed by a $2\times 2$ mass matrix:
\bea
&&{\mathcal M}(1^{++})=2 M_{[cq]}+\frac{\kappa_{c\bar c}+\kappa_{q\bar q}}{2}- \kappa_{c\bar q}-(\kappa_{cq})_{\bar 3} \\
&&{\mathcal M}(1^{+-})=2 M_{[cq]} -\frac{\kappa_{c\bar c}+\kappa_{q\bar q}}{2}+ \left(\begin{array}{cc}
\kappa_{c\bar q}-(\kappa_{cq})_{\bar 3} &\kappa_{q\bar q}-\kappa_{c\bar c}\\ \kappa_{q\bar q}-\kappa_{c\bar c} &-\kappa_{c\bar q}+(\kappa_{cq})_{\bar 3}    \end{array}\right)
\label{masscminus}
\eea
Decays, observed and (not yet?) observed are:
\bea
&& X(3872)\to J/\Psi+\rho/\omega,~D{\bar D^*}\label{xtojpsi}\\
&& Z_c\to J/\Psi~\pi, J/\Psi~\eta~?, \eta_c~\rho ?
\eea

\paragraph*{{\bf Corresponding molecular states}.}
It is interesting to note the prediction of the molecular picture~\cite{zcmol}. In the latter case, one would associate $S$-wave bound states to the $D \bar D^*$ and  $ D^* \bar D^*$ thresholds. One finds also in this case three $J^P=1^+$ states, to wit
\bea
&& C=+1:\; \frac{|D,{\bar D}^*\rangle+|{\bar D}, D^*\rangle}{\sqrt{2}} ;  \; (3876~\text{MeV})   \nonumber \\ 
&& C=-1:\; \frac{|D,{\bar D}^*\rangle-|{\bar D}, D^*\rangle}{\sqrt{2}}  ;  \; (3875~\text{MeV}) \nonumber \\
&& C=-1:\; |D^*,{\bar D}^*\rangle_{J=1}  ;  \; (4017~\text{MeV})
\label{ZDmol}
\eea
where for each state we have given the value of the corresponding threshold for the charged $D$ and $\bar D^*$ meson pairs. Loosely bound charged molecules should differ by few~MeV's.
There are still two states of opposite $C$ approximately degenerate and close to the $X(3872)$ but the third state is at mass higher  than the $X(3872)$ and $Z_c(3900)$ masses.

\section{The unexpected $J^{PC}= 1^{--}$ resonance Y(4260) and its unexpected 
descendants}

The first $Y$-like unexpected charmonium was discovered by Babar in 2006,  in processes with Initial State Radiation~\cite{4260}:
\be
e^+~e^- \to \gamma~(ISR)+Y(4260)\to \gamma+J/\Psi~\pi^+ \pi^-
\label{xprod}
\ee

 Since the photon is emitted by the initial pair, the resonance at 4260 MeV has $J^{PC}=1^{--}$ is interpreted as a $P$-wave tetraquark. The Dalitz plot of the decay shows rather clearly the $f_0(980)$ band, which led us~\cite{Maiani:2005pe} to assume the quark composition:
\be
Y(4260)=[cs][\bar c \bar s]_{{\rm P-wave}},~S_{[cs]}=S_{[\bar c \bar s]}=0
\ee
However, after discovery of the instanton effect, one realizes that the ansatz:
\be
Y(4260)=[cq][\bar c \bar q],~q=u,~d
\ee
would also give rise to the $f_0$ in the final state while fitting better the mass of this particle as an orbital excitation. I shall follow this intepretation.

The final state in (\ref{xprod}) is in line with the de-excitation mode:
\be
 Y(4260)\to Z_c^\pm (3900)+ \pi^\mp
 \ee

\begin{figure}[htb!]
 \begin{minipage}[c]{12cm}
   \centering
   \includegraphics[width=12cm]{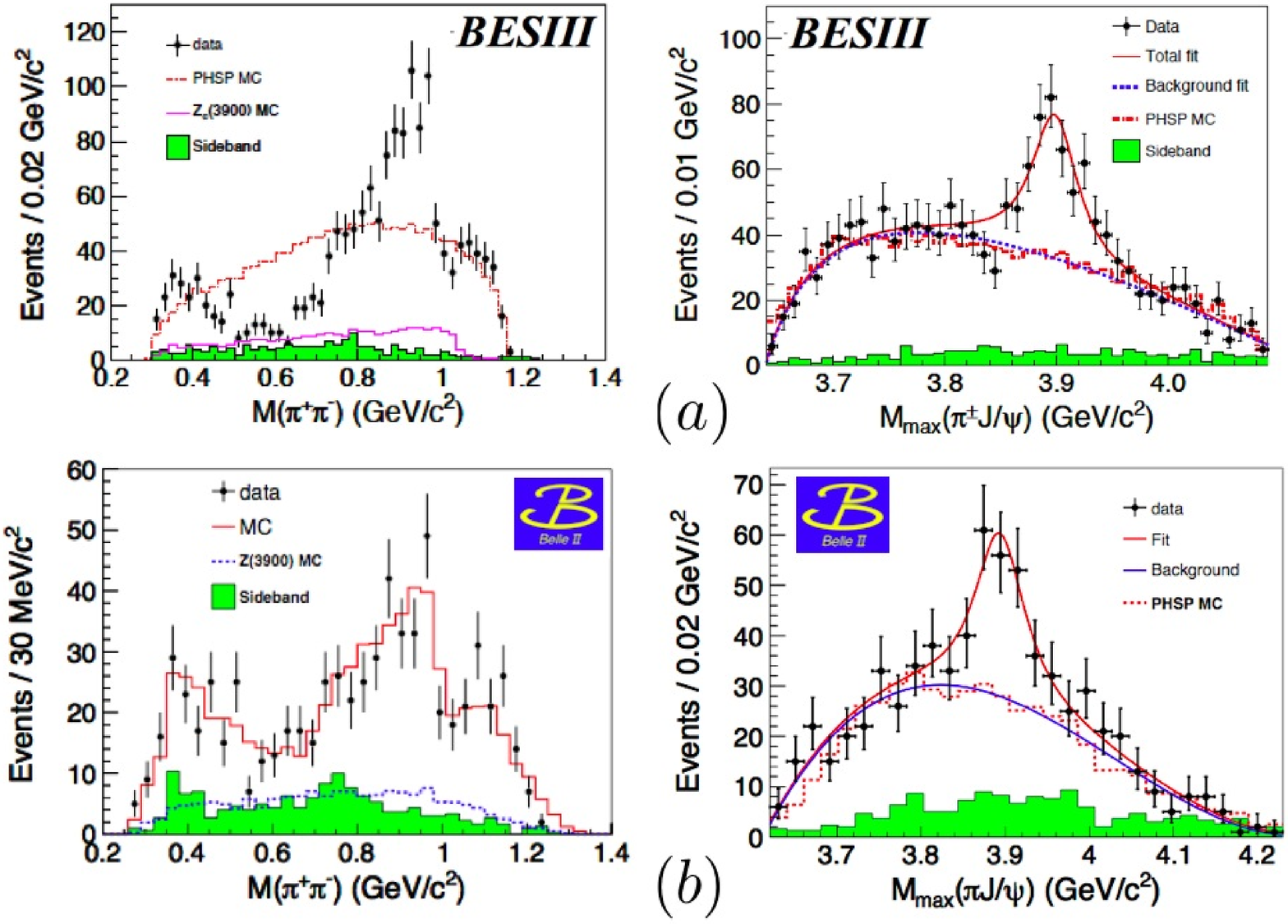}
 \end{minipage}%
 \begin{minipage}[c]{6cm}
  \centering
   \includegraphics[width=6cm]{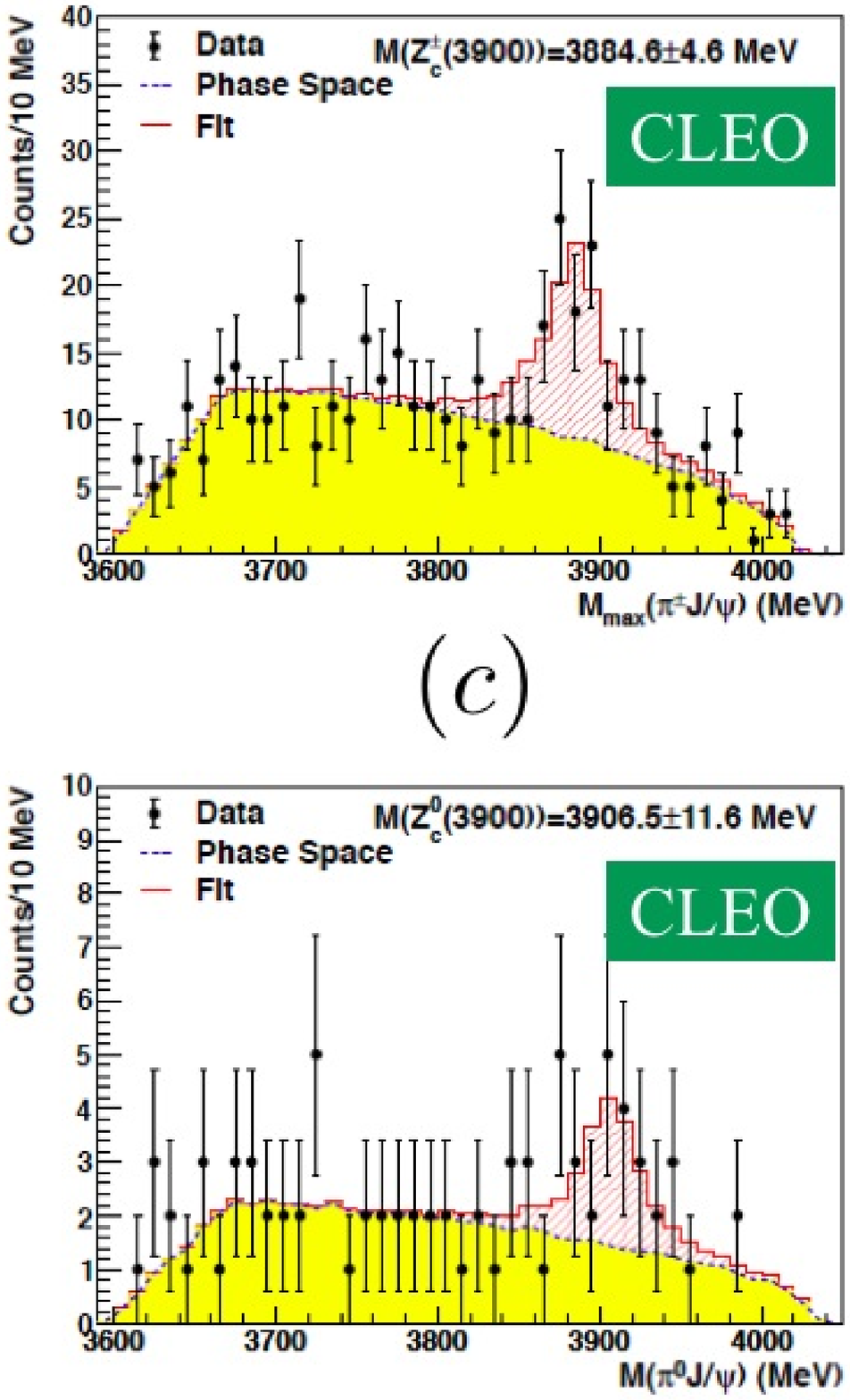}
 \end{minipage}
   \caption{ \label{massproject} \footnotesize  Mass projections of the Dalitz plot of the decay  in (\ref{ydecay}) reported by: (a) BES III~\cite{Ablikim:2013mio}, (b) BELLE~\cite{Liu:2013dau}, (c) CLEO~\cite{Xiao:2013iha}. The distribution for the neutral $Z_c$ reported by CLEO is also shown in (c).}
\end{figure}

BES~III~\cite{Ablikim:2013mio},~BELLE~\cite{Liu:2013dau} and~CLEO~\cite{Xiao:2013iha} have observed the decays:
\bea
&& Y(4260)\to \left\{ \begin{array}{c}  J/\Psi ~f_0\to J/\Psi~\pi^+ \pi^-\\
 Z_c^\pm~\pi^\mp \to J/\Psi~\pi^+ \pi^-\end{array}\right\}
 \label{ydecay}
\eea
\begin{figure}[t]
\begin{center}
\includegraphics[scale=.5]{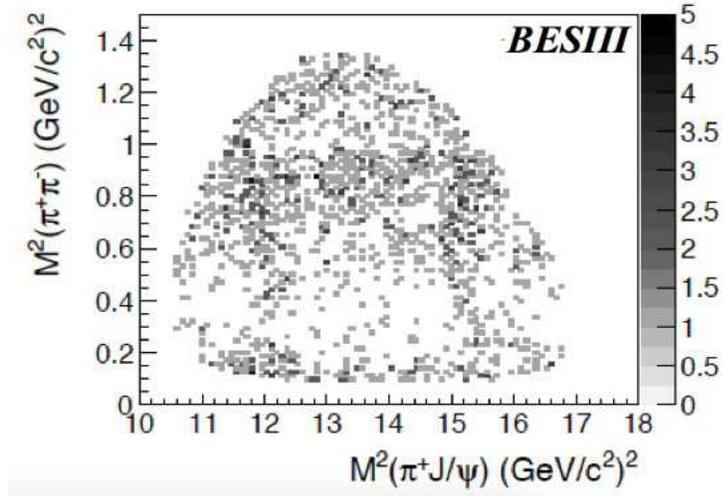}       
\caption{\label{dalplot}  {\footnotesize Dalitz plot of the decay in (\ref{ydecay}) reported by BES III~\cite{Ablikim:2013mio}. The $f_0$ line (horizontal, at $M^2\approx 0.9$) and the $Z_c$ line (vertical, at $M^2\approx 15$), are visible.}}
\end{center}
\end{figure}
Fig.~\ref{dalplot} shows the Dalitz plot reported by BES III, with the visible lines corresponding to the $f_0$ and to the $Z^+_c(3900)$. At a similar mass, CLEO reports a $3\sigma$ signal of a $Z^0_c$ line in the decay:
\be
Y(4260)\to Z_c^0~\pi^0 \to  J/\Psi~\pi^0 \pi^0
\ee
The projected mass spectra are reported in Fig.~\ref{massproject}.

As we have seen, the tetraquark model predicts also second resonance in the same channel, around 3800 MeV, whose possible presence is discussed in~\cite{Faccini:2013lda}. The molecular model, on the other hand, would correlate the second peak to the $D^* {\bar D}^*$ threshold, some 100 MeV above. So, the decay channel (\ref{ydecay}) may provide a significant opportunity to discriminate the two pictures. More precise experiments  are clearly called for.

The situation can be compared with the hidden beauty situation, where two peaks, $Z_b(10610)$ and $Z_b^\prime(10650)$, are clearly seen in the decay of ${\Upsilon}(5S)$: why this difference?

\paragraph*{{\bf Heavy quark spin conservation.}} Spin changing interactions in QCD are inversely proportional to quark masses. Thus it is reasonable to assume that the total spin of the heavy quarks is conserved in strong decays. The value of the total $c\bar c$ spin in the initial state would favour decay modes into $J/\Psi$ ($S_{c\bar c}=1$) or $\eta_c$ ($S_{c\bar c}=0$) accordingly.

It is easy to see that the state in~(\ref{cplus}) has $S_{c\bar c}=1$. In fact, to obtain $C=+1$ one has to have $S_{q\bar q}=S_{c\bar c}$, see eq.~(\ref{c-conj}), and the pair $(S_{q\bar q},S_{c\bar c})=(0,0)$ cannot reproduce the total angular momentum $J=1$. This goes well with the observed decays of $X(3872)$ into $J/\Psi$ in eq.~(\ref{xtojpsi}).

For the states~(\ref{cminus}) we may argue as follows.

The matrices representing the operators $({\vec s}_q\cdot {\vec s}_{\bar q})$ and $({\vec s}_c\cdot {\vec s}_{\bar c})$ in the basis~(\ref{cminus}) are easily found from eq.~(\ref{masscminus}):
\bea
&& 2({\vec s}_q\cdot {\vec s}_{\bar q})=\frac{\delta {\mathcal M}(1^{+-})}{\delta \kappa_{q\bar q}}=-\frac{1}{2} + \sigma_1\notag \\
&& 2({\vec s}_c\cdot {\vec s}_{\bar c})=\frac{\delta {\mathcal M}(1^{+-})}{\delta \kappa_{c\bar c}}=-\frac{1}{2} - \sigma_1
\label{spinmatrix}
\eea

These two matrices are simultaneously diagonalized by two eigenvectors:
\be
v_\pm=\frac{1}{\sqrt{2}}\left(\begin{array}{c}1\\\pm1\end{array}\right)
\ee

The plus sign corresponds to the eigenvalue: $2({\vec s}_q\cdot {\vec s}_{\bar q})=+1/2$, i.e. $S_{q\bar q}=1$, and the minus sign to: $2({\vec s}_q\cdot {\vec s}_{\bar q})=-3/2$, i.e. $S_{q\bar q}=0$. In correspondence, we have for $S_{c\bar c}$ the complementary values: $S_{c\bar c}=0$ and $S_{c\bar c}=1$, as required by the negative Charge Conjugation value of the states~(\ref{cminus}), see again eq.~(\ref{c-conj}).

Denoting by $v_{high}$ the higher mass eigenvector of ${\mathcal M}$, eq.~(\ref{masscminus}), the probability of its decay into $J/\Psi$ is proportional to:
\be
P(J/\Psi)= |<v_{-}|v_{high}>|^2
\ee

The parameters of Tab.~\ref{tab:couplings} lead to a higher eigenvector very close to $v_+$, therefore suppress the decay into $J/\Psi$ vs. the one into $\eta_c$. However this depends critically from the assuptions  used to obtain the spin coupling between the quark and the antiquarks inside different diquarks from the meson couplings. If one allows for a reduction of the overlap function $|\psi(0)|^2$, one can move $v_{high}$ towards $v_{-}$. 

To illustrate the trend, we fix $(\kappa_{cq})_{\bar 3}$ from Tab.~\ref{tab:couplings} and replace $\kappa_{q\bar q}\to r\cdot \kappa_{q\bar q}$, allowing for a reduction factor $r=1.0,0.3, 0.0$ for all quark-antiquark couplings, always fixing the diquark mass from
$X(3872)$. We find the results shown in Tab.~\ref{tab:redfact}.


  \begin{table}[hb]%
\label{redfact}
\centering
\begin{tabular}{|lc|l|clc|lc|c|}
\hline
r && $M_{high}$ && $M_{low}$&& $M_{[cq]}$ && $P(J/\Psi)$ \\
\hline
$1.0$ && $3882$ && $3754$ && $1933$ && $0.0$\\
$0.3$ && $3897$ && $3846$ && $1943$ && $0.12$\\
$0.0$ && $3916$ && $3872$ && $1947$ && $0.50$\\
\hline 
\end{tabular}
\caption{{\footnotesize }
} 
\label{tab:redfact} 
\end{table}

A measurement of the decay rate of $Z_c^\pm \to\eta_c \rho^\pm$ would be extremely informative.

\paragraph*{\bf {Decay modes and widths of Zc(3900)}.}The $Z_c(3900)$ resonance is much broader than $X(3872)$. In Ref.~\cite{Faccini:2013lda}, on the basis of phase space and order of magnitude dimensional arguments, we estimate a total rate of $60$~MeV for the tetraquark model. For molecular models, see~\cite{mol1,mol2}.
 





\section{Critical open issues}

There are a few critical issues which could distinguish between systems bound by QCD forces between colored objects (tetraquark) or by residual forces between colo singlet objects (molecular models in different variations).

The first issue is that of complete isospin multiplets. In the first option, one should see neutral partners of Z and charged partners of X and Y particles. The states observed until now correspond to channels involving $J/\Psi$ + charged pions.  The others require the observation of final states with one or more $\pi^0$s, which may be more diffcult to pin down.

Another issue is isospin conservation, which would give intersting clues on the dynamics of bound states and their decay. Is isospin maximally violated with mixing between $I=0, 1$ states ?

A third issue is the conservation of total heavy quark spin, e.g $S_{c\bar c}$, which is very plausible in QCD.  The control of the decays into $\eta_c$, again with many neutral particles in the final state, is required and it would allow to probe into the structure of the four quark  bound states, wether made by colored or by  color neutral subsystems.

Finally, the existence of exotic states with different angular momentum and parity would be very illustrative of the angular momentum of the total spin of the subsystems: are there $J=2$ or $J=0$ exotic hidden charm or beauty states?

\section{Conclusions}

Exotic mesons, with hidden charm and beauty ARE there, but their nature is still unclear. Correlation of their mass to meson meson thresholds would speak in favor of these mesons to be meson-meson molecules. However, their sizeable production at hadron colliders disfavors extended structures vs. compact structures, with subsystems bound by color forces, of which tetraquarks are working examples.

The search for one additional peak in $Y(4260)$ decays, below or above the $Z_c (3900)$ mass may be crucial to tell molecules from tetraquarks.

The search for the missing partners with respect to electric charge and spin  is called for, in all channels.
This involves, in general, the control of decay modes with $\pi^0$, $\eta$, $\eta_c$, which may be quite a formidable challenge.

All this may take quite some time to be accomplished... and a lot of work at flavor factories and LHC, to keep us busy for a while.

\section*{{\bf NOTE ADDED}} In fall 2013, important new informations have been added by BESIII to the issue of the charged $Z_c$ states, which we analyze briefly here. 
\begin{itemize}
\item A $(D\bar D^*)^\pm$ 
mass peak in $e^+e^- \to \pi D\bar D^*$ is observed~\cite{DD*}, with $M=3884$~MeV and $\Gamma=25$~MeV, consistent within errors with being another decay mode of $Z_c(3900)$ with an estimated rate in $(D\bar D^*)^\pm$ about $6$ times the $\pi~J/\Psi$ rate. 
\item A charged $D^*\bar D^*$ state  with $G=+1$  and presumably $J^P=1^+$ has been observed with $M=4026$~MeV~\cite{D*D*}, some $10$~MeV above the $D^*\bar D^*$ threshold. 
\item The mass of $Z_c(4026)$ is at variance with the tetraquark model prediction discussed here in Eq.~(\ref{cminus}). It requires a change in the spin-spin interaction, suggesting a very small reduction factor, $r$, for the quark-antiquark, spin-spin couplings and a value of the quark-quark coupling, $(\kappa_{cq})_3$, larger than what estimated from the baryon spectrum. As seen in Tab.~\ref{tab:redfact}, a small $r$ was also indicated by the seemingly unsuppressed rate $Z_c(3900) \to J/\Psi  \pi^\pm$. 
\item For $r= 0$, one finds  the two first states in (\ref{cminus}) to be degenerate and the third one at a larger mass, with $\Delta M=+2  (\kappa_{cq})_3$. The spectrum in Fig.~\ref{spectrum2}, in the $r=0$ case, would comprise  one $J^P=2^+$ state and one $0^+$ state at the same mass as the $D^*\bar D^*$ state, and one additional $0^+$ state with $\Delta M=-2(\kappa_{cq})_3$ with respect to $Z_c(3900)$. Similar considerations would apply to the $Z_b$ states.
\item An additional peak in the channel $h_c(1P) \pi^\pm$ is observed at $M=4026$~\cite{hcpi}. This particle would have $G=+1$, similarly to the previous ones, but to be the same particle as the $D^*\bar D^*$ state it should decay in $P$-wave. It is not clear at this moment if this is supported by data. $S$-wave decay would indicate $J^P=1^-$, i.e. a charged $Y$ state.
\end{itemize}  
\section*{Acknowledgements}  
I am indebted to A. Polosa for numerous discussions on the subject of this lecture.  Interesting discussions with F. Close, A. Pilloni, V. Riquer and X. Shen are gratefully acknowledged. I thank Prof. A. Zichichi for supporting my partecipation in the Erice School.


\end{document}